# New Method for Quantitative Determination of Volatile Compounds in Spirit Drinks by Gas Chromatography. Ethanol as Internal Standard.

**Siarhei V. Charapitsa, Anton N. Kavalenka, Nikita V. Kulevich, Nicolai M. Makoed, Arkadzi L. Mazanik, Svetlana N. Sytova**

**Research Institute for Nuclear Problems of Belarusian State University, POB 220089, Gruschevskaya Str., 124, Minsk, Belarus**, Tel/Fax: 375-172-26-2517, <chere@inp.bsu.by>

**Abstract**

The new methodical approach of using ethanol as internal standard in gas chromatographic analysis of volatile compounds in spirit drinks in daily practice of testing laboratories is proposed. This method provides determination of volatile compounds concentrations in spirit drinks directly expressed in milligrams per liter (mg/L) of absolute alcohol according to official methods without measuring of alcohol strength of analyzed sample. The experimental demonstration of this method for determination of volatile compounds in spirit drinks by gas chromatography is described.

**Introduction**

According to the official methods [1-4] the accredited laboratories should determine the following volatile compounds in spirit drinks: acetaldehyde, methyl acetate, ethyl acetate, methanol, 2-propanol, 1-propanol, isobutyl alcohol, n-butanol, isoamyl alcohol. Concentrations of these compounds are expressed in milligrams per liter (mg/L) of absolute alcohol (AA). For determination of impurity concentrations by gas chromatographic methods (GC) the internal standard (IS) method is typically used [1-3, 5-7], with pental-3-ol as IS. Some researchers [5, 8] accomplish determinations by means of the external standard (ES) methods, either experimental simplicity or to avoid the introduction of another source of error with the addition of an internal standard (even though it is well established scientific principle that internal standards tend to increase the precision and accuracy of analytical methods). In order to obtain quantitative values of impurities per liter of absolute alcohol it is also required to measure alcohol strength of analyzed sample [1-4].

It was proposed [9-12] that the main component (solvent) could be used as an internal standard for determination of impurity concentrations. It is now feasible to implement this approach for routine practice of analytical laboratories due to the wide linear dynamic range of modern GC with flame ionization detector (FID), which is generally more than $10^7$. For quantitative determination of specified impurities in spirit drinks, calibration of the chromatographic system includes determination of relative detector response factors for every analyzed compound relative ethanol, which is the major (solvent) in the chromatogram.

**Theoretical background**

The main difference of the proposed method "ethanol as ISTD" from classical method of IS in this case is the following.

In the classical case calibration of chromatograph includes the measuring of relative detector response factors for every analyzed compound relative to IS. Numeric values of these factors $RF_i$ are calculated from chromatographic data for standard solutions with known concentrations of analyzed compounds and may be expressed by the following equation:

$$RF_i = \frac{A_{IS}^{st}}{A_i^{st}} / \frac{C_{IS}^{st}(mg/L)}{C_i^{st}(mg/L)} = \frac{A_{IS}^{st} \cdot C_i^{st}(mg/L)}{A_i^{st} \cdot C_{IS}^{st}(mg/L)}, \qquad (1)$$

where $A_i^{st}$ and $A_{IS}^{st}$ are peak areas of *i*-th compounds and IS respectively; $C_i^{st}(mg/L)$ and $C_{IS}^{st}(mg/L)$ are concentrations of *i*-th compounds and IS respectively expressed in mg per 1 liter of solution.

Concentration of *i*-th sample compound relative to absolute alcohol $C_i$ [mg/L] is expressed by the following formula [1–3]:

$$C_i = RF_i \times \frac{A_i}{A_{IS}} \times C_{IS}(mg/L) \times \frac{100}{Strength}, \qquad (2)$$

where $A_i$ and $A_{IS}$ are the peak areas for *i*-th compound and IS respectively, $C_{IS}(mg/L)$ is concentration of IS, *Strength* is concentration of alcohol in solution expressed in % volume.

In the case of "ethanol as ISTD" the formulas (1) and (2) looks as follows:

$$RF_i(Et\_as\_IS) = \frac{A_{IS}^{st}}{A_i^{st}} / \frac{C_{Et}^{st}(mg/L(AA))}{C_i^{st}(mg/L(AA))} = \frac{A_{IS}^{st} \cdot C_i^{st}(mg/L(AA))}{A_i^{st} \cdot \rho_{Et}}, \qquad (3)$$

where $C_i^{st}(mg/L(AA))$ is concentration of *i*-th compounds expressed in mg per 1 liter of absolute alcohol, $\rho_{Et}$=789300 mg/L is the density of ethanol.

Concentration of *i*-th sample compound relative to absolute alcohol $C_i$ [mg/L(AA)] is expressed by the following formula:

$$C_i = RF_i(Et\_as\_IS) \times \frac{A_i}{A_{Et}} \times \rho_{Et}. \qquad (4)$$

According to (4) we obtain value of *i*-th sample compound concentration directly expressed in mg per 1 liter of absolute alcohol directly without any additional measurement of strength and without of any procedure of IS adding in an analyzed sample.

**Standard and sample preparations**

All individual standard compounds were purchased from Sigma-Fluka-Aldrich (Berlin, Germany). The standard solutions for graduation and sample solutions for researches were prepared by adding of individual standard compounds in ethanol-water mixture (96:4). Ethanol of high grade quality was purchased from Minsk-Kristall Winery and Distillery Plant (Minsk, Belarus).

**Gas Chromatographic conditions**

Analyses were carried out on the gas chromatograph Crystal5000 (JSC SDB Chromatec, Yoshkar-Ola, Russia) equipped with FID, a split/splitless injector, liquid autosampler, Unichrom software (New Analytical Systems Ltd., Minsk, Belarus), capillary column Rt-Wax, 60 m x 0.53 mm, phase thickness 1 µm (Restek, Bellefonte, PA, USA). The oven temperature was: initial isotherm at 75 °C (9 min), raised to 155 °C at rate 7 °C/min with final isotherm of 155 °C (2.6 min). Carrier gas was nitrogen. Gas flow was 2.44 mL/min; injector temperature 160 °C; detector temperature 200 °C; injector volume 0.5 µL and split ratio 1:20. This high split ratio was chosen to achieve good separation between peaks of 2-propanol and ethanol.

**Results and discussion**

Once the gas chromatographic conditions had been optimized the satisfactory separation under these conditions has been achieved. Typical chromatogram of the used standard solutions is presented in Fig. 1-2.

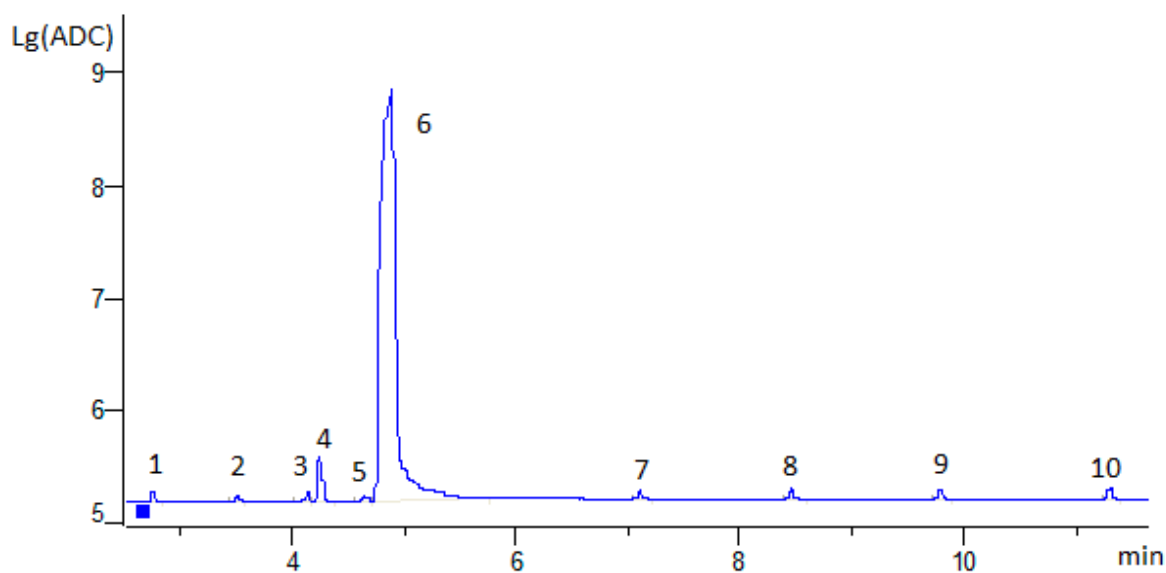

**Fig.1.** Typical chromatogram of standard ethanol-water (40% and 60%) solutions. To show the dominant compound ethanol and another minor compounds synchronously the logarithm scale of response signal is chosen. 1 - acetaldehyde, 2- methyl acetate, 3 - ethyl acetate, 4 - methanol, 5 - 2-propanol, 6 - ethanol, 7 - 1-propanol, 8 - isobutyl alcohol, 9 - n-butanol, 10- isoamyl alcohol.

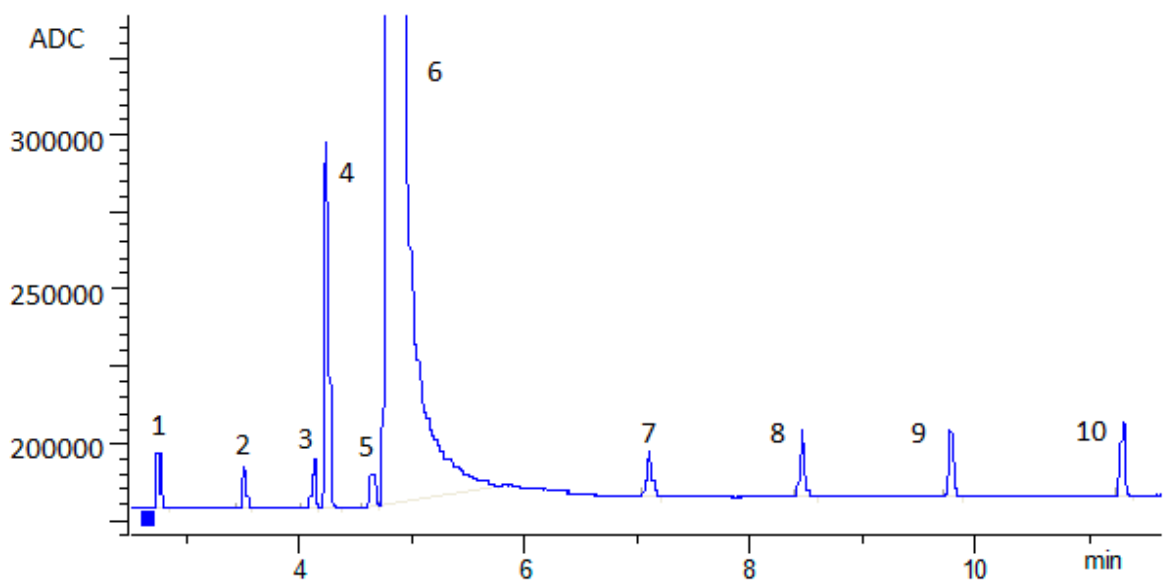

**Fig.2.** The same chromatogram as in Fig. 1, but linear scale of response signal is chosen.

Six standard ethanol-water (96:4) solutions were prepared to generate calibration curves. There were the following levels of volatile compounds concentrations: 13 mg/L, 20 mg/L, 500 mg/L, 1000 mg/L, 5000 mg/L and 20000 mg/L for methanol and 2 mg/L, 5 mg/L, 10 mg/L, 100 mg/L, 500 mg/L and 2000 mg/L for another eight volatile compounds. Every standard solution was injected three times. Analytical characteristics of the obtained calibration curves are presented in Table 1. As can be seen in Table 1, correlation coefficients $R^2$ for all compounds are higher than 0.9996. Detection limits were determined by analysis of low level standards. The detection limits are between 0.235 mg/L for isobutyl alcohol and 0.394 mg/L for methanol.

**Table 1.** Analytical characteristics of the obtained calibration graphs of volatile compounds in standard ethanol-water (96:4) solutions.

| Compound | Linear range (mg/L) | Slope | Correlation coefficient ($R^2$) | LOD* (mg/L) |
|---|---|---|---|---|
| acetaldehyde | 2.24 - 1990 | 1.559 | 0.9996 | 0.289 |
| methyl acetate | 2.09 - 2000 | 1.517 | 0.9997 | 0.333 |
| ethyl acetate | 2.20 - 2094 | 1.247 | 0.9998 | 0.322 |
| methanol | 13.0 - 20045 | 1.377 | 0.9999 | 0.394 |
| 2-propanol | 3.74 - 2033 | 0.914 | 0.9998 | 0.319 |
| 1-propanol | 1.99 - 2094 | 0.809 | 0.9998 | 0.262 |
| isobutyl alcohol | 2.23 - 2000 | 0.674 | 0.9998 | 0.235 |
| n-butanol | 1.98 - 2000 | 0.737 | 0.9998 | 0.267 |
| isoamyl alcohol | 2.18 - 2073 | 0.681 | 0.9999 | 0.276 |
| * limit of detection (LOD) | | | | |

In order to study accuracy of the proposed methodical approach in the case of large ranges of volatile compounds concentrations 6 – 20000 mg/L for methanol and 1 – 2000 mg/L for another eight volatile compounds reference ethanol-water solutions were prepared with known concentrations of volatile compounds. Every reference solution was injected 30 (15 × 2) times. The repeatability in the worst case for lower concentrations 1 mg/L did not exceed 3.6 %. The obtained experimental results are presented in Table 2.

**Table 2.** Experimental measured concentrations of volatile compounds in reference ethanol-water (96:4) solutions.

| Compound | Known concentration of i-th compound in standard, (mg/L) | Concentration measured by IS method, (mg/L) | Relative discrepancy, % |
|---|---|---|---|
| acetaldehyde | 1.158 | 1.129 | -2.50 |
| | 5.137 | 5.182 | 0.88 |
| | 10.11 | 9.921 | 1.87 |
| | 99.64 | 93.86 | -5.80 |
| | 497.6 | 481.1 | -3.32 |
| | 1989 | 2037 | 2.42 |
| methyl acetate | 1.000 | 1.005 | 0.50 |
| | 5.000 | 5.121 | 2.42 |
| | 10.00 | 9.905 | -0.95 |
| | 100.0 | 96.35 | -3.65 |
| | 500.0 | 484.9 | -3.02 |
| | 2000 | 2042 | 2.10 |
| ethyl acetate | 1.047 | 1.072 | 2.39 |
| | 5.234 | 5.374 | 2.67 |
| | 10.47 | 10.45 | -0.19 |
| | 104.7 | 102.0 | -2.58 |
| | 523.4 | 512.2 | -2.14 |
| | 2093 | 2115 | 1.02 |
| methanol | 5.975 | 6.044 | 1.15 |
| | 53.07 | 53.51 | 0.83 |
| | 103.2 | 102.97 | -0.22 |
| | 1005 | 988.1 | -1.68 |
| | 5013 | 4987 | -0.52 |
| | 20045 | 20118 | 0.36 |
| 2-propanol | 2.636 | 2.645 | 0.34 |
| | 6.698 | 6.754 | 0.84 |
| | 11.78 | 11.77 | -0.08 |
| | 103.0 | 101.0 | -2.13 |
| | 509.0 | 503.2 | -1.22 |
| | 2033 | 2047 | 0.69 |
| 1-propanol | 1.047 | 0.997 | -4.78 |
| | 5.234 | 5.223 | -0.21 |
| | 10.21 | 10.23 | 0.20 |
| | 103.2 | 100.2 | -3.10 |
| | 523.4 | 513.6 | -1.87 |
| | 2094 | 2125 | 1.51 |
| isobutyl alcohol | 1.000 | 0.971 | -2.90 |
| | 5.000 | 5.033 | 0.66 |
| | 10.00 | 9.82 | -1.80 |
| | 100.0 | 97.7 | -2.30 |
| | 500.0 | 491 | -1.80 |
| | 2000 | 2032 | 1.60 |
| n-butanol | 1.000 | 0.991 | -0.90 |
| | 5.000 | 5.061 | 1.22 |
| | 10.00 | 9.89 | -1.10 |
| | 100.0 | 97.10 | -2.90 |
| | 500.0 | 491.0 | -1.80 |
| | 2000 | 2036 | 1.80 |
| isoamyl alcohol | 1.036 | 1.003 | -3.19 |
| | 5.182 | 5.169 | -0.25 |
| | 10.37 | 10.21 | -1.54 |
| | 104.0 | 101.0 | -2.60 |
| | 518.0 | 510.0 | -1.58 |
| | 2073 | 2110 | 1.78 |

The concentrations of analyzed volatile compounds calculated according to IS method are expressed directly in milligrams per liter (mg/L) of absolute alcohol. It is not necessary to do additional measurements of alcohol strength in this case and potential error in value of ethanol concentration is eliminated from resulting formula. There is significant simplification of total measurement procedure.

Verification of method stability against dilution of testing samples was been done by the next way. Three reference ethanol-water solutions were analyzed after dilution with water in ratio 1:1 and 1:3. The obtained results are presented in Table 3.

Estimated volume of relative accuracy for proposed method followed from analysis of the obtained experimental in accordance with [13] does not exceed 11 %.

**Table 3.** Experimental measured concentrations of volatile compounds in reference ethanol-water (96:4) solutions after dilution with water in ratio 1:1 and 1:3.

| Compound | Known concentration of i-th compound in standard, (mg/L) | Measured concentration after dilution 1:1, (mg/L) | Relative discrepancy, % | Measured concentration after dilution 1:3, (mg/L) | Relative Discrepancy, % |
|---|---|---|---|---|---|
| acetaldehyde | 10.11 | 10.34 | 2.2 | 10.50 | 3.8 |
|  | 99.64 | 97.28 | -2.4 | 97.40 | -7.3 |
|  | 497.6 | 483.3 | -2.9 | 473.1 | -4.9 |
| methyl acetate | 10.00 | 10.25 | 2.5 | 9.78 | -2.2 |
|  | 100.0 | 92.76 | -7.2 | 89.17 | -10.8 |
|  | 500.0 | 463.7 | -7.3 | 452.4 | -9.5 |
| ethyl acetate | 10.47 | 10.18 | -2.8 | 10.63 | 1.6 |
|  | 104.7 | 100.0 | -4.5 | 95.46 | -8.8 |
|  | 523.4 | 489.6 | -6.5 | 477.3 | -8.8 |
| methanol | 103.2 | 97.99 | -5.0 | 95.18 | -7.8 |
|  | 1005 | 921.9 | -8.3 | 904.1 | -10.0 |
|  | 5013 | 4654 | -7.2 | 4514 | -9.9 |
| 2-propanol | 11.80 | 11.63 | -1.2 | 10.56 | -10.4 |
|  | 103.2 | 97.86 | -5.2 | 93.13 | -9.7 |
|  | 509.4 | 479.5 | -5.9 | 463.7 | -9.0 |
| 1-propanol | 10.21 | 10.36 | 1.5 | 10.01 | -1.9 |
|  | 102.1 | 98.00 | -4.0 | 96.23 | -5.7 |
|  | 523.4 | 482.9 | -5.8 | 483.2 | -7.7 |
| isobutyl alcohol | 10.00 | 10.42 | 4.2 | 10.35 | 3.5 |
|  | 100.0 | 96.87 | -3.1 | 94.32 | -5.7 |
|  | 500.0 | 480.1 | -4.0 | 471.5 | -5.7 |
| n-butanol | 10.00 | 10.17 | 1.7 | 9.98 | -0.2 |
|  | 100.0 | 97.02 | -3.0 | 95.21 | -4.8 |
|  | 500.0 | 482.9 | -3.4 | 475.2 | -5.0 |
| isoamyl alcohol | 10.37 | 11.28 | 8.8 | 10.35 | -0.2 |
|  | 103.7 | 103.0 | -0.6 | 99.52 | -4.0 |
|  | 518.2 | 509.0 | -1.8 | 500.6 | -3.4 |

**Conclusion**

The goal of this work is to show possibility of the new methodical approach of using ethanol as internal standard in gas chromatographic analysis of volatile compounds in spirit drinks in daily practice of analytical and testing laboratories.

In the next paper we will present experimental results concerned with the further verification and validation of the method "ethanol as ISTD" by comparison of all three methods cited above. Results were obtained in the Laboratory of Analytical Research of the Institute for Nuclear Problems (INP) of Belarusian State University and in the Control Laboratory of Bobruisk Hydrolysis Plant from Belarus (BHP). There is a good coincidence between these results.

Thousands of testing laboratories all over the world carry out gas chromatographic analysis of volatile compounds in spirit drinks day-and-night. They may test this approach in their real practice. It is important to note that there is no need to perform any additional measurements. This method could be tested while performing current measurements with existing instrumentation. Calculations could be done in parallel by three different methods: using traditional IS method [*1–3*], ES method [*4*] and using ethanol as IS.

Manual and algorithm of the proposed methodical approach "ethanol as ISTD" can be found here http://www.inp.bsu.by/labs/lar/eis.html .

**Acknowledgements**

The authors would like to thank the RUE "Minsk-Kristall" Winery and Distillery Plant for supplying the ethanol of high grade quality, the Institute of Physical Organic Chemistry of the National Academy of Sciences of Belarus for purification of reagents and New Analytical Systems Ltd. for instrumentation support. Special thanks from the authors to Dr. Vadims Bartkevics from the Institute of Food Safety, Animal Health and Environmental "BIOR" of Latvia and Jaap de Zeeuw from Restek Inc for their invaluable guidance and encouragement while doing this project. We thanks Prof. Nataliya I. Zayats from Department of Physical and Chemical methods of certification of products, Belarus State Technology University for assistance in metrological treatment of experimental data We are very grateful to the Head of the Control Laboratory of Bobruisk Hydrolysis Plant Olga P. Leskovets and her analysts for large work on validation of the method "ethanol as ISTD".